\documentstyle[11pt,paspconf,epsf]{article}
\def\spose#1{\hbox to 0pt{#1\hss}}
\def\simlt{\mathrel{\spose{\lower 3pt\hbox{$\mathchar"218$}}
     \raise 2.0pt\hbox{$\mathchar"13C$}}}
\def\simgt{\mathrel{\spose{\lower 3pt\hbox{$\mathchar"218$}}
     \raise 2.0pt\hbox{$\mathchar"13E$}}}

\def\edcomment#1{\iffalse\marginpar{\raggedright\sl#1\/}\else\relax\fi}
\reversemarginpar

\begin{document}
\title{Spectral properties of galaxies in the Stromlo-APM redshift
survey: clues on the local star-forming galaxies} 
\author{Laurence Tresse$^1$, Steve J. Maddox$^1$ and Jon Loveday$^2$} 
\affil{$^1$Institute of Astronomy, Madingley Road, Cambridge CB3 0HA, UK\\
$^2$University of Chicago, 5640 S Ellis Ave, Chicago, IL 60637, USA} 

\begin{abstract}
We analyse emission-line properties of the $b_j$-selected
Strom\-lo-APM spectra ($\langle z \rangle = 0.05$). Because this is a
representative sample, we can study the global spectral properties of
the local galaxy population.  We classify spectra according to their
H$\alpha$ emission, which is closely related to massive star
formation.  This study gives a comparative local point for analysis of
more distant surveys.  We show that in the local universe, faint,
small galaxies are dominated by star formation activity, while bright,
large galaxies are more quiescent.  Obviously this picture of the
local universe is quite different from the distant one, where bright
galaxies appear to show a rapidly-increasing activity back in time.
\end{abstract}

\section{The sample}
We measured emission lines for $1671$ spectra in the Stromlo-APM
(SAPM) redshift survey of galaxies with $ 15 \ge b_J \ge 17.15$ mag,
covering 4300 square degrees of the southern sky. Galaxies are at
$z<0.13$, and $\langle z \rangle = 0.051$ (Loveday et al. 1996). These
galaxies have been selected randomly at a rate of 1 in 20 from
automated plate measurements (APM) scans (Maddox et al. 1990).
Spectra were obtained using the Dual Beam Spectrograph equipped with 4
CCDs on the ANU 2.3m telescope.  The wavelength coverage is
3700$-$5000 \AA\ in the blue, and 6300$-$7600 \AA\ in the red.  The
dispersion is typically 1 \AA\ per pixel, and using a slit of width
8\arcsec\, the spectral resolution is 5 \AA.  We checked that our
equivalent widths (EW) are not dependent on the area of each galaxy
covered by the slit, which is $\sim 50$\% on average.  Consequently,
our results are comparable to integrated spectra of nearby galaxies,
or to distant galaxies (Tresse et al. 1998).
 
\section{Emission and absorption H$\alpha$ line galaxies} 
We find that 62\% of galaxies are H$\alpha$ emission-line galaxies
(ELG) with $EW(H\alpha) \\ \ge2$ \AA. The remaining are classified as
absorption-line galaxies (ALG). The sample is divided into equal
numbers of ELG and ALG at $EW(H\alpha)\simeq 5$ \AA. EW are not
corrected for stellar absorption.  When comparing surveys, percentages
of ELGs have to be taken with extreme care, since the spectral
resolution, the aperture, the magnitude limit, the passband selection,
etc.  lead to very different percentages.  The fraction of ELG depends
strongly on the absolute magnitudes, M($b_j$), and the size of
galaxies (Figure~1). From $-23$ to $-18$ mag, or from size 10,000 to
100 kpc$^2$, it increases by a factor of $2$ or 4, respectively.  ELG
are more common than ALG in intrinsically faint galaxies, and small
systems.  This agrees with the general picture where dwarves (faint
and small) are more actively forming stars, than their bright
counterparts in the local universe.  This is a reflection of the
difference between ELG and ALG luminosity functions (Loveday et
al. 1998).  It is also consistent with the ``peak'' of SFR observed at
$z\sim1-2$, since the overall luminosity density is dominated by
bright galaxies, which formed most of their stars much earlier than
dwarves, which are actively forming stars today.

\begin{figure}
\plotfiddle{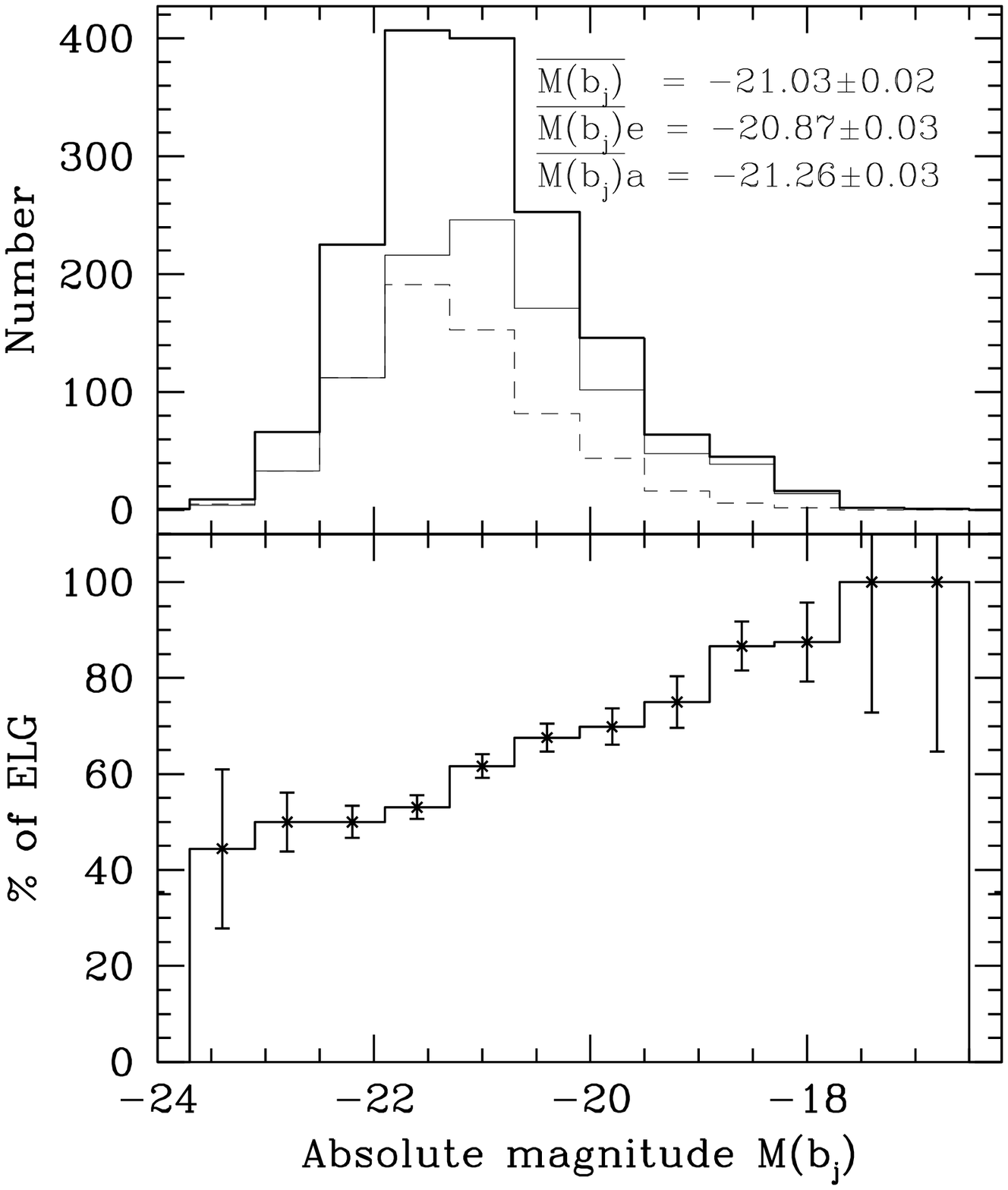}{2cm}{0}{35}{35}{-205}{-180} 
\plotfiddle{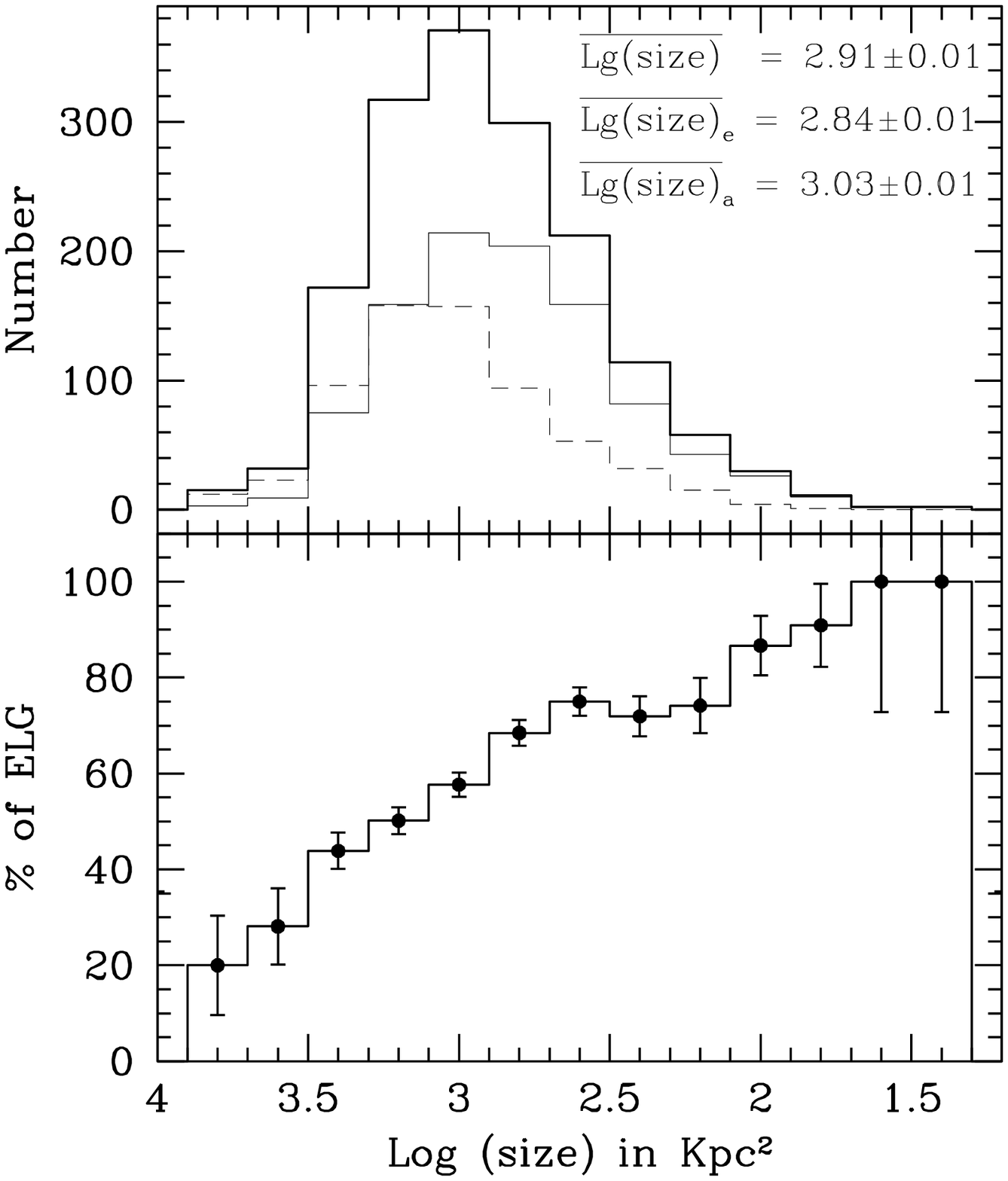}{2cm}{0}{35}{35}{-10}{-110}
\vspace*{2cm}
\caption{\small Top-left panel: N(M(b$_j$)) distributions. Bottom-left
panel: Percentage of ELG as a function of M(b$_j$). Top-right panel:
Projected size distributions. Bottom-right panel: Percentage of ELG as
a function of size, where size is defined as area brighter than
$\mu(b_j)=25$ mag/arcsec. Key to histograms: whole sample of $1671$
galaxies (thick solid line), ELG (thin solid line), ALG (dashed
line).}
\end{figure}

\section{H$\alpha$ emission-line galaxies}
The mean and the median EW(H$\alpha$) ($>2$ \AA) are respectively 19
and 15 \AA\ in the SAPM.  The H$\alpha$ line has the advantage that it
is directly correlated to the ``instantaneous'' star formation rate
(SFR), or the time scale since the last burst. Whereas
[O~II]$\lambda$3727 line depends on the metallicity of the galaxy, and
on the hardness of the global ionizing spectrum.  EW(H$\alpha$) is the
ratio of the flux originating from massive, young stars (OB type) in
H~II regions, over the flux from the old stellar population emitting
in the rest-frame $R$ passband, which forms the continuum at
H$\alpha$. In Figure~2, we show that the fraction of faint ELG
($M(b_j)> -21$) increases as a function of EW. From $EW(H\alpha)=10$
to $60$ \AA\ (which corresponds to a SFR increasing by a factor $\sim$
5), the fraction increases by a factor $\sim 2$. Another way to look
at this is to consider the fraction of high EW ELG ($EW(H\alpha)>15$
\AA), which increases at fainter M(b$_j$). 
Since B magnitudes are correlated with H$\alpha$ luminosities, the
fainter a galaxy, the smaller its H$\alpha$ luminosity is (Tresse \&
Maddox 1998, and also these Proceedings). So, faint galaxies with high
EW(H$\alpha$), means they are blue galaxies. This corresponds also to
the strong correlation between colors and EW(H$\alpha$+[N~II])
(Kennicutt \& Kent 1983).  They have a young stellar population, so
blue colors, high H$\alpha$ fluxes, and less old stars.  Massive,
large galaxies, or bright galaxies do not have the strongest
EW(H$\alpha$). The fraction of young stars to old stars is smaller
than for faint galaxies, leading to a red continuum, and smaller EW on
average.  So they are older systems, and less numerous --- but they
are still contributing to most of the luminosity density in deep
surveys.  It is now well established that the faint end of the 
luminosity function of $B$-selected surveys is dominated by actively
star-forming galaxies.  This is what we find also in the SAPM (Loveday
et al. 1998, $\alpha \approx -1.2$ and $-1.3$ for $EW(H\alpha)>5 $ and
19 \AA).  We note that the overall SAPM luminosity function at
$\langle z\rangle=0.05$ has a shallower slope than the one from the
preliminary 2dF data ($\langle z \rangle = 0.1$), which were selected
from the same parent catalog as the SAPM (see Maddox, these
Proceedings).

\begin{figure}
\plotfiddle{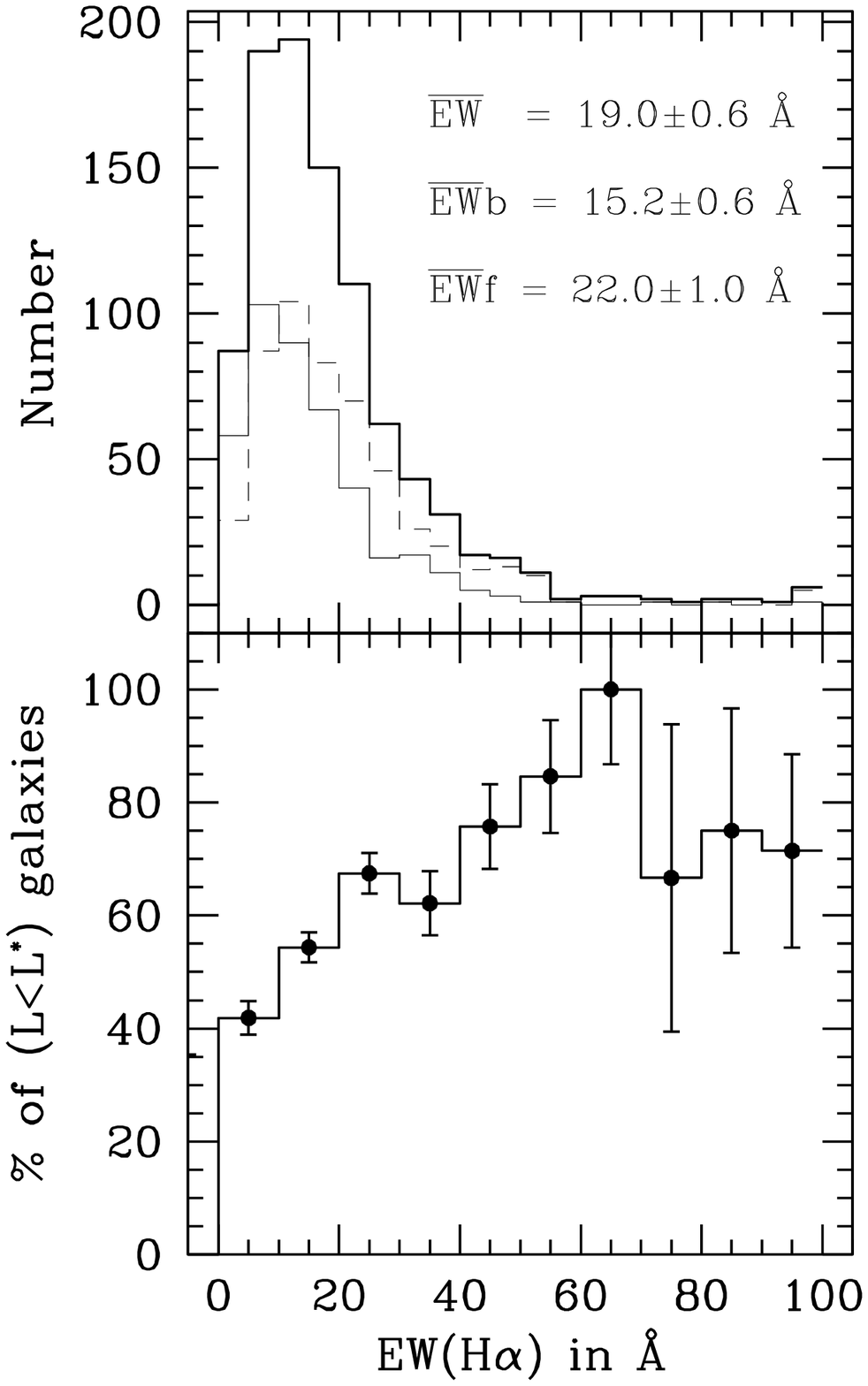}{2cm}{0}{50}{35}{-225}{-185} 
\plotfiddle{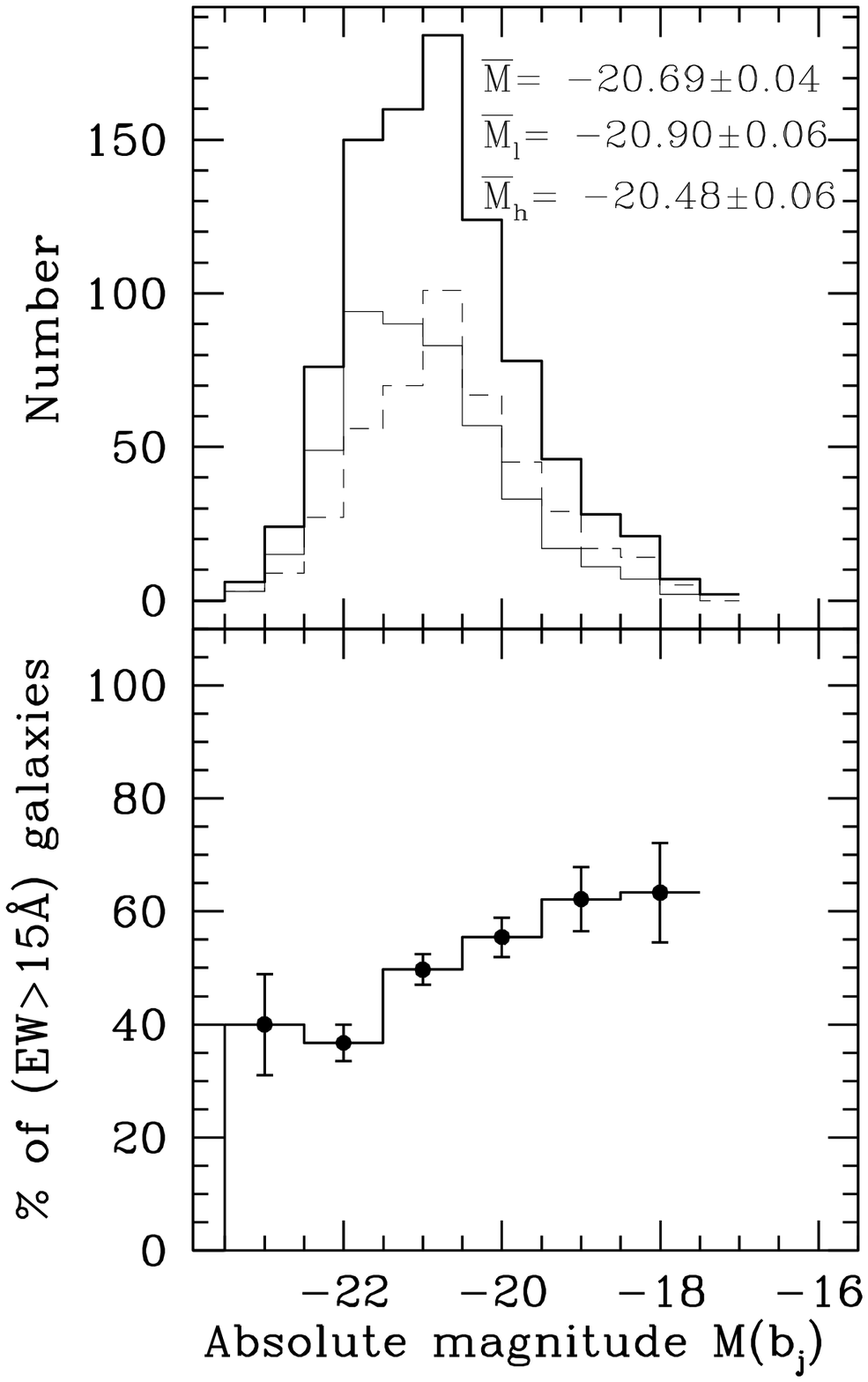}{2cm}{0}{50}{35}{-30}{-115}
\vspace*{2.5cm}
\caption{\small Top-left panel: EW(H$\alpha$) distributions for all
ELG (933) (thick line), for bright ELG (416), M(b$_j$)$\leq -21$ (thin
line), for faint ELG (517), M(b$_j$)$> -21$ (dashed line). Bottom-left
panel: Percentage of faint galaxies as a function of
EW(H$\alpha$). Top-right panel: M(b$_j$) distributions for all 933 ELG
(thick line), for low EW ELG (471), EW$\leq$ 15 \AA, (thin line), for
high EW ELG (462), EW$>$ 15 \AA, (dashed line). Bottom-right panel:
Percentage of high EW ELG as a function of M(b$_j$). }
\end{figure}

\section{H$\alpha$ and [O~II] correlations}   
The correlation between EW(H$\alpha$) and EW([O~II]) has been
established by Kennicutt 1992 (K92), using a selection of galaxies
covering all spectral types.  However, K92's selection is not a truly
representative statistical sample of the local universe, unlike our
sample.  In Figure~3, we find $EW([O~II]) = 0.68\ EW(H\alpha)$ (K92:
0.61), $EW([O~II]) = 0.47\ EW(H\alpha + [N~II])$ (K92: 0.40). This is
consistent with K92's result given the large scatter observed in both
samples.  Our scatter is not due to poor signal-to-noise, but reflects
certainly different stellar, dust and metal contents sampled.  We
emphasize that EW measured from the overall galaxy content are likely
to be affected by dust. Indeed, recombination emission lines originate
from dusty H~II regions, where hot stars (OB type) are formed, whereas
the continuum at [O~II] or H$\alpha$ comes from long-lived stars
sitting in less, or non obscured regions.  EW([O~II]) is more affected
by the dust than EW(H$\alpha$). Hence, the absolute EW relation should
be in average slightly higher.

\begin{figure}
\plotfiddle{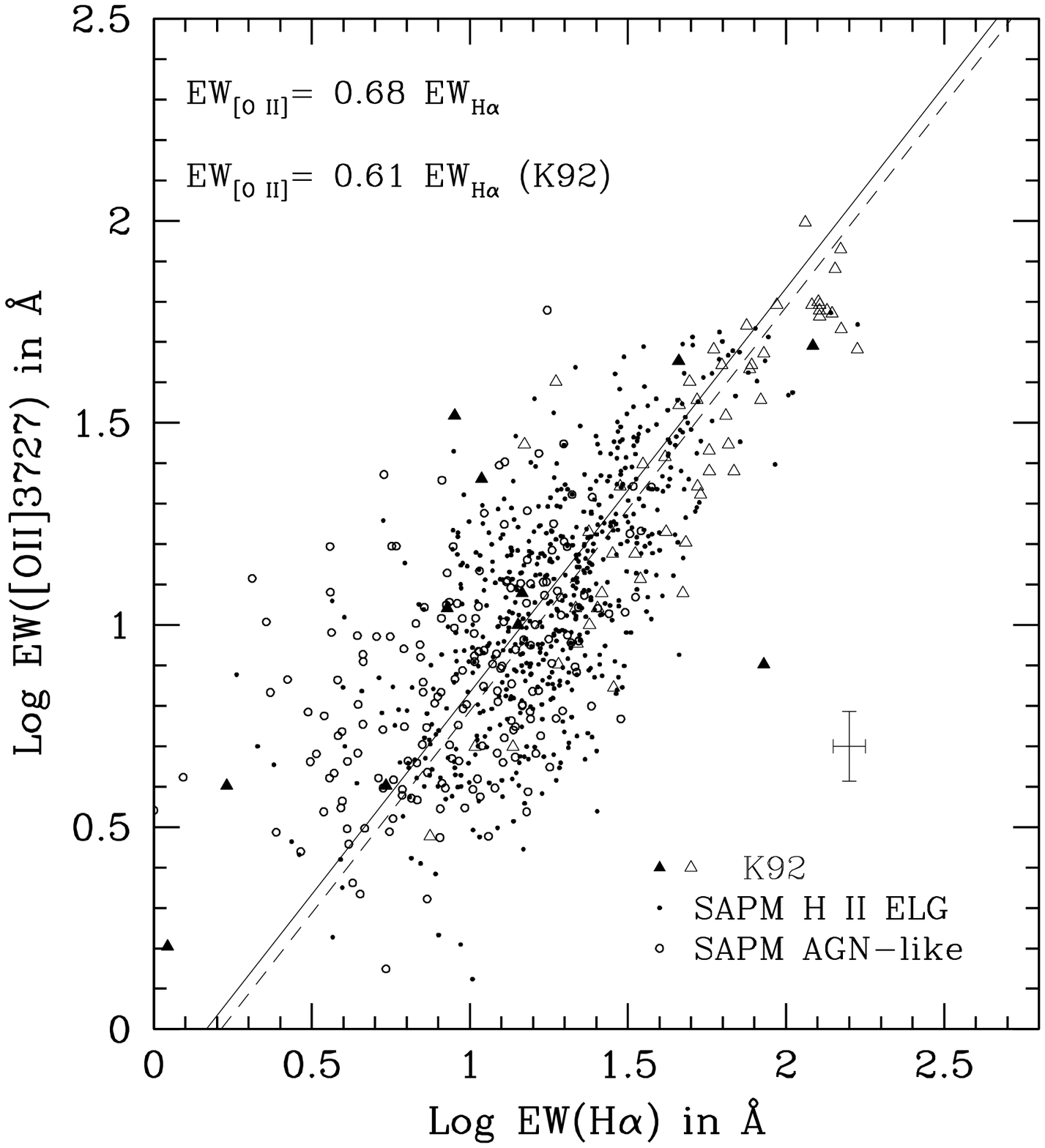}{2cm}{0}{33}{33}{-210}{-170}
\plotfiddle{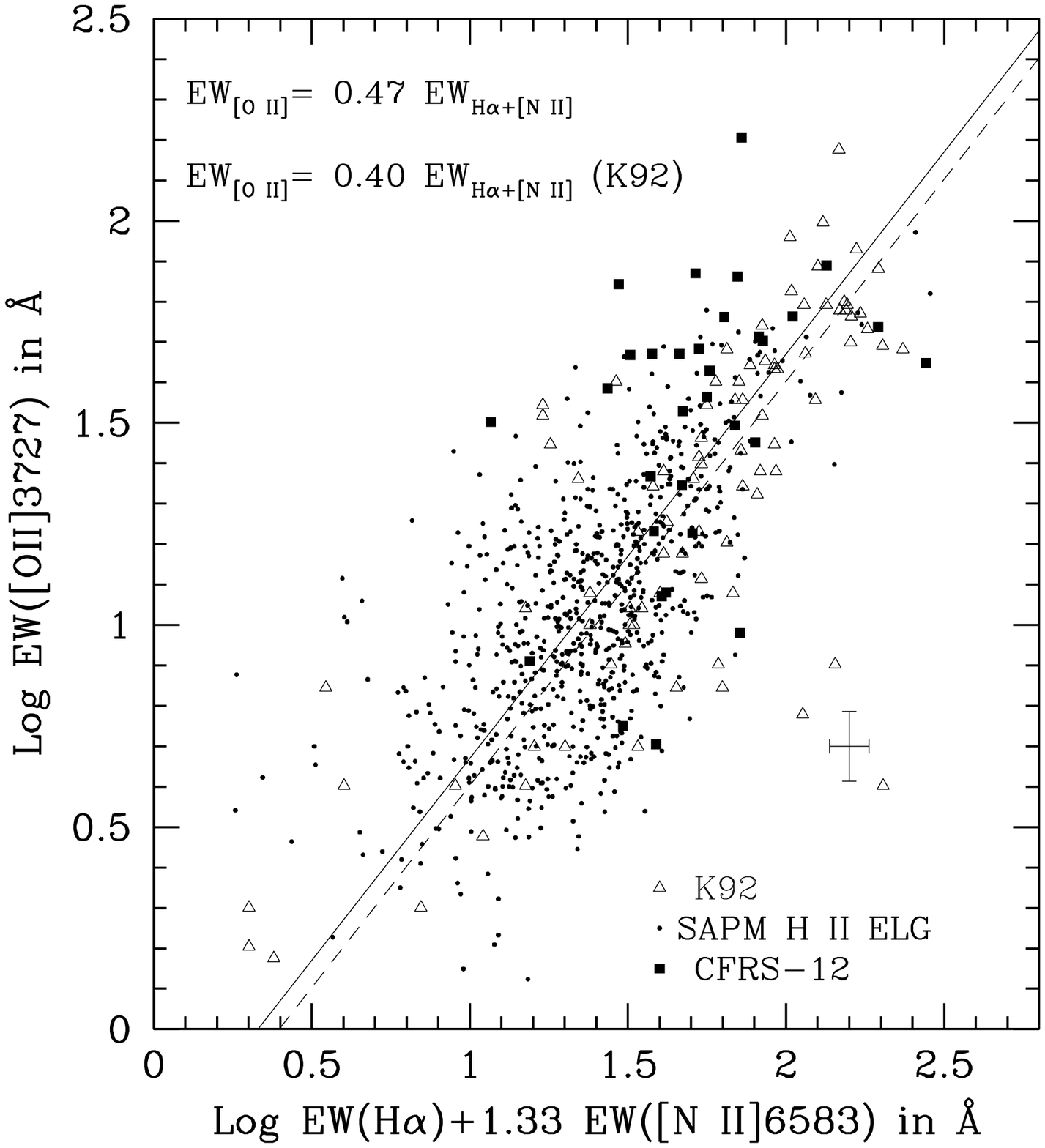}{2cm}{0}{33}{33}{-15}{-100}
\vspace*{2cm}
\caption{\small Left-panel: EW([O~II]) versus EW(H$\alpha$). Open
symbols are AGN-like spectra. The plain line is our relation, the
dashed one the K92 one. Right-panel: EW([O~II]) versus
EW(H$\alpha$) +1.33 EW([N II]6583).  The dashed line shows the K92
relation EW(OII)$=$0.4 EW(H$\alpha$+NII). The square symbols are
CFRS-12 data at $\langle z \rangle = 0.2 $ (Tresse et al. 1996).}
\end{figure}

\end{document}